\documentclass[prb,twocolumn,aps,showpacs,footinbib,10pt,numerical,superscriptaddress]{revtex4-2}
\usepackage[english]{babel}
\usepackage[utf8]{inputenc}
\usepackage{amsmath}
\usepackage{amsfonts}
\usepackage{amssymb}
\usepackage{graphicx}
\usepackage{lmodern}
\usepackage{color}
\usepackage{enumerate}
\usepackage{mathtools}
\usepackage[pdfstartview=FitH,colorlinks=true,citecolor=blue,linkcolor=blue,urlcolor=blue]{hyperref}
\usepackage[normalem]{ulem}
\usepackage{float}
\usepackage[caption=false]{subfig}
\usepackage{enumitem}
\usepackage{verbatim}
\usepackage[]{placeins}
\usepackage{lipsum}

\begin{document}

\newcommand{\beginsupplement}{%
        \setcounter{table}{0}
        \renewcommand{\thetable}{S\arabic{table}}%
        \setcounter{figure}{0}
        \renewcommand{\thefigure}{S\arabic{figure}}%
     }

\title{Waiting time distributions in Quantum spin hall based heterostructures}
\author{F.~Schulz}
\affiliation{Department of Physics, University of Basel, Klingelbergstrasse 82, CH-4056 Basel, Switzerland}

\author{ D.~Chevallier}
\affiliation{Bleximo Corp., 701 Heinz Ave, Berkeley, CA 94710, USA}

\author{ M.~Albert}
\affiliation{Universit\'e C\^ote d'Azur, CNRS, Institut de Physique de Nice, 06560 Valbonne, France}

\date{\today}

\begin{abstract}
For the distinction of the Andreev bound states and Majorana bound states, we study the waiting time distributions (WTDs) for heterostructures, based on one dimensional edge states of a two dimensional topological insulators (TI) in combination with an proximitized s-wave superconductor (SC) and an applied magnetic field. We show for the time reversal symmetric (TRS) situation of a Josephson junction details of the WTD. This includes different transport processes, different numbers of Andreev bound states and the phase difference of the SC. We further consider a Zeeman field in the normal part of the junctions revealing novel features in the WTD along the phase transition between trivial bound states and Majorana bound states. We finally discuss clear signatures to discriminate between them.\\
\vspace{1cm}
\end{abstract}

\maketitle

The quest of Majorana fermionic states in condensed-matter physics, have been the subject of a strong interest during the past decades, in particular because of their exotic properties, such as non-Abelian statistics, that open the way to using them for quantum computation~\cite{beenaker2013, dassarma}. Among many proposals to create them, heterostructures based on two dimensional TIs are one interesting lead of research. Previous investigations on the topologically non-trivial bulk band inversion in TIs result in quasi one-dimensional TRS protected counter propagating edge states~\cite{BHZ,koenig1,koenig2}. These spin polarized states in combination with an $s$-wave superconductor (SC) and a magnetic field can be used to engineer Majorana Bound States~(MBSs) within different geometrical setups~\cite{fu2008,fu2009,alicea2012,schrade2015}. Induced superconductivity in the edge states of TIs has experimentally been realized in quantum well setups~\cite{HgTe1,HgTe2,HgTe3,HgTe4,InAs1,InAs3,InAs4}, or alternatively on thin-layer TIs~\cite{BiSe1,BiSe2,WTe21,WTe22,WTe23}. The combination with a magnetic field for such systems results in a zero energy mode in the conductance, potentially allowing the characterization of a MBS~\cite{BiSe3}. The realization of topological superconductivity has also been realized on semiconducting nanowires~\cite{nanowire1,nanowire2,nanowire3}. However, usual conductance measurements to identify fingerprints of MBS for such topological setups remain elusive, such that we are motivated for alternative measurements. One possibility would be to resort to the Waiting Time Distributions~(WTDs) \cite{Brandes2008,WTD1,WTD1b,WTD2,WTD3,WTD3b,WTD5,WTD6,WTD7,WTD8,WTD9,WTD10,WTD11}, namely the distribution of time delay between the detection of two consecutive charge carriers, which has been shown to reveal traces of MBS in such non-trivial systems~\cite{WTD3b,WTD7,WTD10}.

Necessary techniques for the counting of single particles have come very precise, allowing a novel measurement for the processes of specific scattering events~\cite{Singleelectron1,Singleelectron2,Singleelectron3,Singleelectron4,Singleelectron5,Singleelectron6,Singleelectron7,Flindt2009,Ubbelohde2012,Basset2012}. The resulting WTD in turn provides statistics including signatures that may manifest a clear distinction between topologically trivial and non trivial states within the system.  WTDs in semiconducting hybrid systems have recently been studied, providing distinct features of a one dimensional p-wave SC in comparison to an s-wave SC~\cite{WTD3b}. Further theoretical investigations have shown results for entangled electrons on a SC interface for MBSs~\cite{WTD7,WTD10}.

\begin{figure}
\includegraphics[width=1\linewidth]{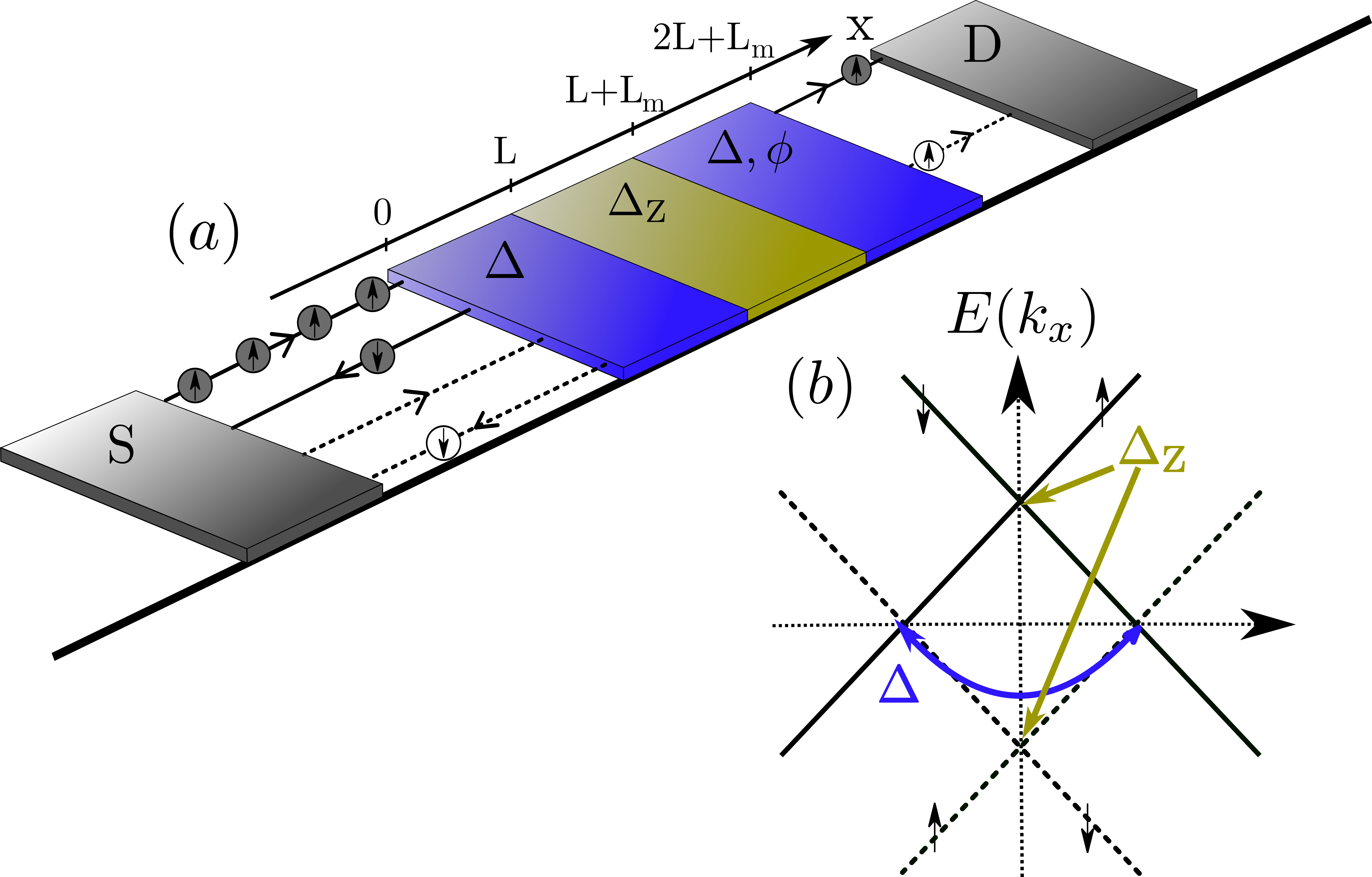}
\caption{(a) Schematic of the setup. The scattering area, containing two superconducting (blue) regions enclosing a ferromagnet (yellow) are connected to source and drain electrodes. Further indicated are the possible scattering states of an incoming spin up electron, which may be reflected as electron/hole or be transmitted as electron/hole. (b) Energy spectra with the effective paring amplitudes in the scattering regions.}
\label{fig.:setup}
\end{figure}
 
In this paper we study WTDs in materials based on topological hybrid junctions as presented on Fig. \ref{fig.:setup}. We combine the WTD as a statistical characterization tool to detect several transport effects that appear within such topological junctions. Starting with a simple application on almost perfectly Andreev reflected electrons, we continue with more complicated situations, such as the inclusion of several transmitting resonances. Finally, by including a magnetic field, we create topological protected MBSs, which we measure and discuss along their distinct signatures in the WTD along the phase transition.  

The article is setup as follows, in Sec.~\ref{sec.:sectionII} we introduce the model that is used for the underlying N-S-F-S-N junction. In Sec.~\ref{sec.:sectionIII} we recap the substance of WTDs. The results are presented in Sec.~\ref{sec.:sectionIV}, where we elaborate WTDs for specific types of hybrid quantum conductors, closely related to the Fu and Kane structure~\cite{fu2009}. We study in details the WTDs for both, ABS and MBS and identify signatures to distinguish them. Conclusion and outlook are given in Sec.~\ref{sec.:sectionV} and several technical details are
available in Appendices.

\section{Model}\label{sec.:sectionII}
The heterostructure we consider consists of helical counter propagating edge states of a Quantum spin hall insulator, proximity coupled to a spatially restricted $s$-wave superconductor and a magnetic field [see Fig.~\ref{fig.:setup} (a)]. Similar setups have been studied in the literature~\cite{fu2008,fu2009, Trauzettel2014,  Trauzettel2018b,Schulz2020}. In the basis $\Psi=(\psi_\uparrow^{},\psi_\downarrow^{},\psi_\uparrow^\dagger,\psi_\downarrow^\dagger)$, the Hamiltonian of the  system under consideration is the following
\begin{equation}\label{eq.:1}
H=\left( \begin{matrix} H_0  & H_\text{SC}^* \\
H_\text{SC} & -H_0^* \end{matrix} \right) ,
\end{equation}
where $H_0=H_\text{TI}+H_\text{Z}$. The Hamiltonian $H_\text{TI}=v_\text{F}k_x\sigma_1-\mu$ is the kinetic term of the edge states, with Fermi-velocity $v_\text{F}$ ($\hbar=1$), momentum $k_x=-i\partial_x$ and chemical potential $\mu$, while $H_\text{Z}=\sigma_1\Delta_\text{Z}(x)$ is the Zeeman field pointing in $x$-direction. Furthermore, $H_\text{SC}=-i\Delta(x)\sigma_2$ includes the superconductor, which is assumed to be grounded, and the Pauli matrices $\sigma$ ($\tau$) act in spin (Nambu) space. The spatial arrangements of $\Delta$ and $\Delta_\text{Z}$ are chosen to have a ferromagnetic region enclosed by two superconducting regions, such that $\Delta (x)=\Delta [\Theta (x)-\Theta(x-L)]+\Delta e^{i\phi}[\Theta (x-L-L_m)-\Theta(x-2L-L_m)]$ and $\Delta_\text{Z}(x)=\Delta_\text{Z}[\Theta (x-L)-\Theta(x-L-L_m)]$. The pairing mechanisms open a gap in the spectrum of the edge states [see schematic in Fig.~\ref{fig.:setup} (b)], which is either at the Dirac-point~(DP)~($\Delta_\text{Z}$) or at the Fermi level crossing~($\Delta$). In addition we also allow the possibility to add a magnetic flux $\phi$ to the system [see Fig.~\ref{fig.:setup} (a)].

We check the characterization of different states appearing in the heterostructure, by first presenting the scattering features as a basis to identify distinct signatures in the WTD. Whereby we use the seminal work of Blonder, Tinkham and Klapwijk (BTK) to calculate the corresponding scattering coefficients~\cite{BTK}. For our interests we study two cases, first, no Zeeman field is applied (trivial phase) and the middle region hosts freely propagating states, and second, a non vanishing Zeeman field (topological phase above a certain treshold) couples the two spin species and opens a gap in the middle region, resulting in the emergence of MBSs on the interfaces of the superconducting and magnetic regions~\cite{fu2009}. We assume in this work a rather long junction ($L_m>v_F/\Delta$), such that for $\Delta_\text{Z}=0$ multiple bound states are well defined within the superconducting gap $E<\Delta$ [see Fig.~\ref{fig.:coef1} (a)]. Calculating the wavefunction with the continuity conditions on the interfaces at $x=0,L,L+L_m$ and $x=2L+L_m$ results in the necessary transport coefficients. With a preserved TRS an incoming right moving electron from the source is ($\Delta_\text{Z}=0$) protected from backscattering, such that there are only two non vanishing processes. Namely, the co-tunneling of an electron ($t_e$) form source to drain and the local Andreev-reflection ($r_h$) on the SC interface at $x=0$. The breaking of TRS in turn then also allows normal electron reflection ($r_e$), including a spin flip, and the transmission of holes ($t_h$). While the gap closing and reopening is usually used as an indicator of the topological phase transition, we use only the normal electron reflection signature, accompanied by its probability at zero energy from zero to one. Details are presented in App.~\ref{sec.:App-Coefficients}. 
Based on those energy dependent coefficients, we are highly interested in the corresponding WTDs, which we introduce in the next section.

\section{Waiting Time Distribution}\label{sec.:sectionIII}
 For the sake of clarity, we recap in this section, the formalism used for the calculation of WTDs~\cite{Brandes2008,WTD1,WTD1b,WTD2} as well as some well established results that will serve as reference. The WTD $\mathcal{W}(\tau)$ denotes the probability distribution for the time delay $\tau$ between the detection of two consecutive charge carriers. They can be electrons or holes for instance. This quantities gives precise and transparent informations about correlations in a transport process. In general, it is customary to calculate it from the Idle Time Probability (ITP) $\Pi ( \tau)$, namely the probability to detect zero particles during a time interval $\tau$. For stationary processes these two distributions are connected by the following expression
\begin{eqnarray}\label{eq.:WTD}
\mathcal{W}(\tau)=\langle \tau \rangle\frac{\partial^2\Pi (\tau)}{\partial \tau^2}\text{.}
\end{eqnarray}
For non-interacting systems, the ITP is given by the determinant formula \cite{WTD1b}
\begin{equation}\label{eq.:ITP}
\Pi(\tau)=\text{det}\big( 1-Q_\tau \big)\text{,}
\end{equation}
where $Q_\tau$ is a projector over the time window $\tau$ whose expression depends on the detection scheme (measurement of consecutives electrons, holes, ...) and the scattering matrix of the system. Explicit formulae are given below for the processes of interest. In addition, the mean waiting time is given by
\begin{equation}
\langle \tau \rangle=-\left.\frac{\partial\Pi (\tau)}{\partial \tau}\right|^{-1} _{\tau=0}.
\end{equation}

To continue, we specify a detection procedure for the WTD. Thus, we calculate the ITP with the projection of a single particle scattering state on an discrete voltage window around the Fermi level. In the range of the applied voltage $V$, the linear energy spectrum is discretized into $N$ compartments with wave vectors $k=\frac{n}{N} \frac{eV}{\hbar v_F}$. A stationary process is then reached by the limit $N\rightarrow\infty$. We label the matrix elements of $Q_\tau$ with the four possible scattering processes as\\
\begin{eqnarray}
\big[Q^{t_e}_\tau \big]_{\text{nm}}&&\approx \frac{\kappa t_e(\kappa n)t_e^*(\kappa m)}{2\pi}K_\tau [\kappa (n-m)],\label{eq.:5}\\
\big[Q^{t_h}_\tau \big]_{\text{nm}}&&\approx \frac{\kappa t_h(\kappa n)t_h^*(\kappa m)}{2\pi}K_\tau [\kappa (n-m)],\label{eq.:6}\\
\big[Q^{r_e}_\tau \big]_{\text{nm}}&&\approx \frac{\kappa r_e(\kappa n)r_e^*(\kappa m)}{2\pi}K_\tau [\kappa (n-m)],\label{eq.:7}\\
\big[Q^{r_h}_\tau \big]_{\text{nm}}&&\approx \frac{\kappa r_h(\kappa n)r_h^*(\kappa m)}{2\pi}K_\tau [\kappa (n-m)]\text{,}\label{eq.:8}
\end{eqnarray}
were $\kappa$ results from the discrete energy compartment to $\frac{1}{N} \frac{eV}{\hbar v_F}$ in the measured energy window $eV$ and the kernel is given by
\begin{equation}
K_\tau[\kappa (n-m)]= \frac{2e^{-i\kappa (n-m)\frac{v_F\tau}{ 2}}\sin (\kappa (n-m)\frac{v_F\tau}{ 2})}{\kappa (n-m)}.
\end{equation}
Thus, we can measure the distribution of waiting times for outgoing electrons or holes from source to drain with Eq.~\eqref{eq.:5} or Eq.~\eqref{eq.:6}. Similar, with the assumption of a grounded superconductor, we calculate the WTDs of reflected electrons Eq.~\eqref{eq.:7} or holes Eq.~\eqref{eq.:8}. The corresponding scattering coefficients, $t_{e/h}$ and $r_{e/h}$, are in general strongly energy dependent and can be seen as an energy filter within the transport window, which will be described later in the text. With this framework we are able to calculate the energy-dependent ITP from the determinant of the matrix in Eq.~\eqref{eq.:ITP}, needed for the WTD [Eq.~\eqref{eq.:WTD}]. Depending on the point of interests, we focus in the next sections on specific scattering processes and their features appearing within the designated WTD.\\

Before going further in the analysis of our model, we first recall some important results that will be useful to understand our work. A reference case is the one of a single quantum channel subjected to a voltage $eV$ in the presence of an energy independent barrier of transmission $T$~\cite{WTD2}. If we look at the WTD of electrons transmitted through the scatterer, it displays a crossover from a Wigner-Dyson distribution at perfect transmission to an exponential at low transmission. Indeed, for a perfectly transmitting channel, in which the scattering state can be seen as a train of non-interacting fermions, the average waiting time in which electrons are separated is given by $\bar{\tau}=h/eV$. This separation is due to the Pauli exclusion and the statistical distribution is approximated by a Wigner-Dyson surmise.

\begin{equation}\label{WD_WTD}
  \mathcal W(\tau)=\frac{32}{\overline{\tau}^3\pi^2} \tau^2\,\exp\left[-\frac{4}{\pi}\frac{\tau^2}{\overline{\tau}^2}\right].
\end{equation}

This shape is easily understood by a mapping between one dimensional fermions to random matrices of the Gaussian Unitary Ensemble \cite{WTD2}. Another important indicator of fermionic statistics is the fact that the WTD vanishes at $\tau=0$. As $T$ is reduced, the WTD spreads to longer waiting times and develops oscillations similar to Friedel oscillations with period $h/eV$. Close to pinch-off ($T\ll 1$) WTD approaches an exponential distribution which is the signature of uncorrelated events. Indeed, in that case the detection time of scattered electrons is very long and conclusively detected electrons are no longer correlated. In this situation, the mean waiting time is simply given by $\langle \tau \rangle = \bar{\tau}/T$, while the corresponding current reads $e/ \langle \tau \rangle=\frac{e^2}{h}VT$, well known as the Landauer formula~\cite{buettiker200}.

\section{Results}\label{sec.:sectionIV}
We present in this section the WTDs, found for two different situations. First, we consider the setup in Fig.~\ref{fig.:setup} (a) without a Zeeman field ($\Delta_\text{Z}=0$), hosting ABS, and second with a Zeeman field lifting the Kramers degeneracy at the DP in the normal region. In the later situation MBS appear on the two interfaces between the superconducting and magnetic regions. Our goal is to discuss the different effects between the two types of bound states (ABS vs. MBS) in terms of their WTDs and finally presenting a possible distinction between the two.

\subsection{Andreev bound states: $\Delta_\text{Z}=0$}\label{sec.:sectionIVa}
We begin with the study of a TRS Josephson-junction by considering the system without a Zeeman field ($\Delta_\text{Z}=0$). Following the BTK scattering formalism~\cite{BTK}, we calculate the scattering probabilities for an incoming electron form the source (see App.~\ref{sec.:App-Coefficients}) with boundary conditions at $x=0, L ,L+L_m$ and $x=2L+L_m$. 

\begin{figure*} 
  \includegraphics[width=\textwidth,height=4cm]{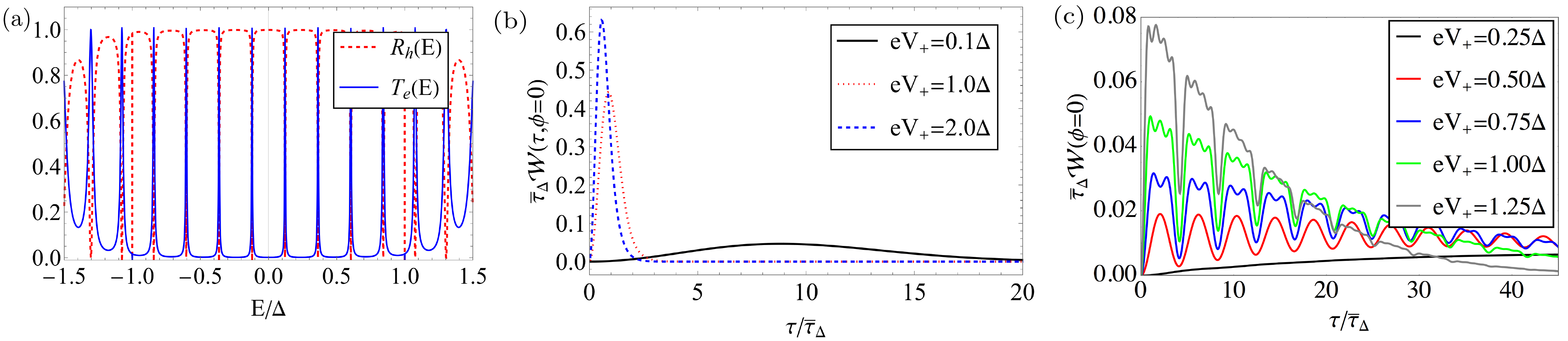}
  \caption{(Color online) \textbf{(a)} Energy dependence for the scattering coefficients $T_e(E)$ (orange) and $R_h(E)$ (black dotted) for $\phi=0$ and $\Delta_\text{Z}=0$. Resonances reflect ABS in the normal region. \textbf{(b)} WTD between holes (Andreev reflected electrons with $R_h(E)$) flowing back to the source for the indicated energy windows referring to (a). \textbf{(c)} WTD for transmitted electrons flowing to the drain. Each distribution covers an increasing number of resonances. The parameters are chosen to be $L=2/\Delta$, $L_m=6L$, $\Delta_\text{Z}=0$.}
 \label{fig.:coef1}
\end{figure*}
To relate the scattering coefficients to the respective WTDs in detail, we first present in Fig.~\ref{fig.:coef1} (a) the energy dependence (at $\phi=0$) of the two coefficients $T_e(E)=|t_e|^2$ (orange) and $R_h(E)=|r_h|^2$ (black dotted), namely the probability that an incoming electron is transmitted or reflected as a hole.
There, we find resonances with a energy difference $\Delta E\propto v_F/L_m$. The transmission probability $T_e(E)$ arises due to the structure of the junction, which can be seen as a Fabry-P\'erot interferometer with the two SC regions acting as barriers. After the transmission of an electron through the first barrier, multiple Andreev reflections within the region enclosed by the two SCs allow those resonances to appear, and are known as Andreev-Bound states. The shape of the peaks can be modulated by the choice of the SC regions, such that a rather short region ($L<\zeta_{SC}$, with the SC coherence length $\zeta_{SC}=v_{F,s}/\Delta$) broadens the spectra. The length of the normal region~($L_m$) modifies the the energy distance (and the number of states) of states within the SC gap. The Andreev-reflection $R_h(E)$ shows dips at the corresponding resonances, which follows directly from the normalization condition, $T_e(E)+R_h(E)=1$.
Next, we evaluate the WTD [Eq.~\eqref{eq.:WTD}] for these processes and present in Fig.~\ref{fig.:coef1} (b) the WTD, were an incoming electron gets Andreev reflected for several voltage windows ($V_+ >0$).
For small voltages (black line) the reflection of a hole is unity and in analogy to a perfectly reflecting channel. Conclusively, the WTD of reflected holes is well aproximated by a Wigner-Dyson distribution, given by Eq. (\ref{WD_WTD}). Note that the indicated reference time is chosen by $\bar{\tau}_\Delta=h/\Delta$, while the WTDs have their maximum around the mean waiting time $\tau=\bar{\tau}=h/eV$. With increasing voltage (red dotted and blue dashed lines), the WTD remains in a Wigner-Dyson distribution, since the dips in the reflection coefficient [see Fig.~\ref{fig.:coef1} (a)] are negligibly small and the channel can be seen as a perfectly reflecting one. Since the scattering coefficient $R_h(E)$ tends to zero for $E\gg\Delta$, a further increase of the voltage would result in an effective smaller reflection within the full window, thus the WTD would evolve towards an exponential distribution (not shown). 

More interesting are WTDs between transmitted electron for the resonances in the scattering coefficient $T_e(E)$. For a better understanding, we choose several $V_+$, such that we cover with each step one more resonance peak of the electron transmission [see black resonances in Fig~\ref{fig.:coef1} (a)]. We highlight the results in Fig.~\ref{fig.:coef1} (c) by presenting the WTDs for several numbers of the transmitting resonances. In analogy to the Fabry-P\'erot interferometry, we find for two resonances (red line) oscillations in the WTD with a period dictated by the energy difference of the two states~\cite{WTD2}. For an increasing number of channels we find on top of those oscillations additional oscillations. Those smaller oscillations in principle allow a counting of the bound states within the applied voltage window $V_+$ and are interpreted as interferences between different paths in energy space \cite{WTD2}. We find that the overall shape of the WTD changes abruptly as an additional resonance is included in the voltage window which can be readily understood in terms of transport through a finite number of resonances as described analyticaly in App.~\ref{sec.:App-Rateeq.}. 

We conclude this section by discussing the flux dependence of the scattering coefficients and the WTDs in the absence of a Zeeman field. The phase dependence of the electron transmission [for details see App.~\ref{sec.:App-Coefficients} Fig.~\ref{fig.:coeff2} (a)] is due to the spin momentum locking and the assumption of a right moving, incoming electron only symmetric in energy $[T_e(E)=T_e(-E)]$ for $\phi=n\pi$, with integer $n\in \mathbb{Z}_0$. Thus, depending on the applied voltage, the flux can either keep the number of included states fixed (voltage is exactly in the middle of two states), or changes it by $\pm1$ (voltage is above/below the middle of two states). Note, since we are in the long junction limit the transmission resonances have a linear dependency and conclusively the phase does not change the period of oscillations. The WTD for Andreev reflected holes stays almost unaffected for a change in $\phi$, as discussed before.

So far we have studied the situation of a time-reversal symmetric system and have not specifically considered the zero-energy state at $\phi=\pi$. For a better understanding, we will then break the TRS and determine the characteristic features in the WTDs.

\subsection{Majorana bound states: $\Delta_\text{Z}\neq 0$}\label{sec.:sectionIVb}

\begin{figure}[h]
\centering
\includegraphics[width=0.90\linewidth]{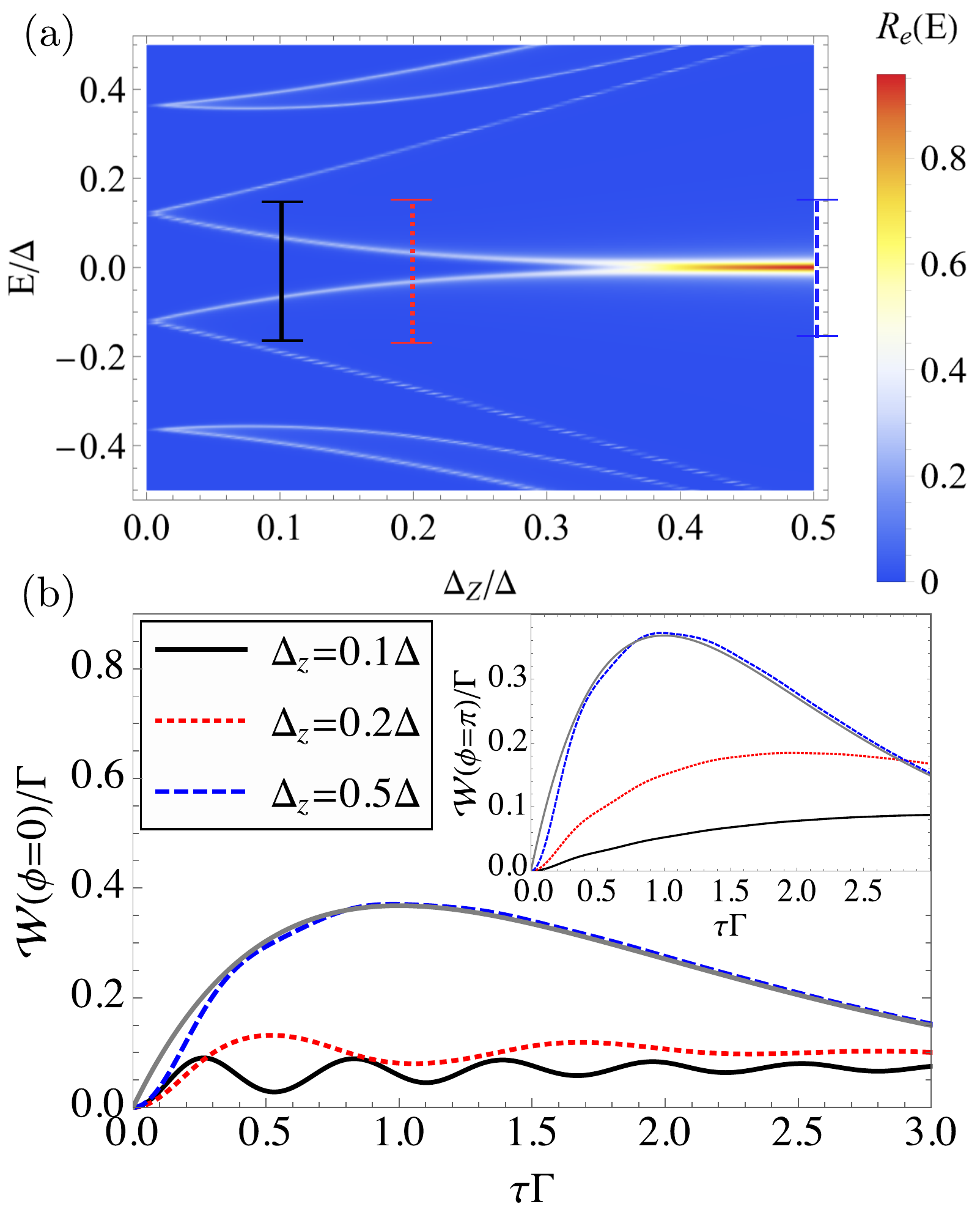}
\caption{(a) Scattering coefficient $R_e(E)$ depending on $E$ and $\Delta_\text{Z}$ at $\phi=0$. The colored bars indicate the measured energy window used in (b), where the respected WTDs are presented. The grey line denotes the analytical result for the rate equation (see App.~\ref{sec.:App-RateeqSD}). The inset shows the results for $\phi=\pi$. The rest of the parameters are those from Fig.~\ref{fig.:coef1} and $\Gamma=0.077\Delta$.}
\label{fig.:coeff3}
\end{figure}

In this section we elaborate the results of the WTD for the transition of ABS to MBS, localized in between or on the interfaces between the ferromagnetic and superconducting regions. The inclusion of a Zeeman field breaks TRS, such that in the scattering process normal electron reflection $R_e(E)$ (with flip) is the major indicator for a phase transition. Furthermore, since both transmission probabilities ($t_e, t_h$) start to vanish with increasing $\Delta_\text{Z}$, we calculate the WTDs for the channels of normal reflected electrons. As before, we first present the electron reflection $R_e(E)$ for an increasing Zeeman field in Fig.~\ref{fig.:coeff3} (a) at energies within the SC gap and without phase-difference $\phi=0$.
A small increase of the magnetic field $\Delta_\text{Z}$ lifts the Kramers degeneracy at the DP and splits the spin degenerated states into their spin up/down components, such that the TRS is broken and normal electron reflection is allowed. The splitting is most pronounced for states at lower energy (at the DP, note that $\mu_{TI}=0$). A stronger rise of the Zeeman field leads to the appearance of the non trivial MBSs, where all other states get pushed to higher energies, while the zero energy anomaly pushes the amplitude of $R_e(0)$ to unity. Note that the criterion for a phase transition in this type of setups dependents on the effective length scales of the superconducting and magnetic regions~\cite{Trauzettel2018}.
For the corresponding signal in the WTD we choose an energy window of $-0.15\Delta$ to $0.15\Delta$ for three values of $\Delta_\text{Z}$ [see the bars in Fig.\ref{fig.:coeff3} (a)]. The resulting WTDs for $\phi=0$ are shown in Fig.\ref{fig.:coeff3} (b).
There, we find for $\Delta_\text{Z}=0.1\Delta$ (black line) an oscillatory behavior in the WTD, since the applied voltage window includes the two lowest energy states [see the corresponding black line in Fig.\ref{fig.:coeff3} (a)]. The period of the oscillations is proportional to the energy distance of the two states and can be described by two shifted Lorenzian resonances (see App.\ref{sec.:App-Rateeq.}). 

The increase of the Zeeman field to $\Delta_\text{Z}=0.2\Delta$ [red dotted line in Fig.\ref{fig.:coeff3} (a) and (b)] still covers the two states with a smaller splitting of the states and conclusively induces a slower period of the oscillations in the WTD. Most strikingly, in the topological phase at $\Delta_\text{Z}=0.5\Delta$ [blue dashed line in Fig.\ref{fig.:coeff3} (a) and (b)]  the oscillatory behavior of the WTD fully disappears and transformed to the distribution of a single channel with a single resonance (see App.~\ref{sec.:App-RateeqSD}), given by $W(\tau)=\Gamma^2\tau \exp (-\Gamma \tau)$ (see gray line) with a maximum at $\tau=1/\Gamma$, while $\Gamma=2\gamma\pi$ is proportional to the Full width at half maximum ($2\gamma$) of the resonance.  Thus, for $\phi=0$ the phase-transition from the non-topological to the topological situation changes the WTD from oscillatory to non oscillatory behavior, which can be used as an alternative detection and distinction of ABSs and MBSs. 
Furthermore, we present in the inset of Fig.~\ref{fig.:coeff3} (b) the WTDs for the same values of $\Delta_\text{Z}$ at $\phi=\pi$ (see Fig.~\ref{fig.:coeff2} (b-d) in App.~\ref{sec.:App-Coefficients} for details of the reflection coefficients). There, the previous oscillations at $\Delta_\text{Z}=0.1$ and $0.2\Delta$ are barely visible, since the electron reflection amplitudes of the higher energy states are negligibly small due to the lifted spin degeneracy of the ABSs at $E \neq 0$. The exceptional point of the phase transition at $R_e(0,\phi=\pi)=|\tanh \left(L_m \Delta _Z\right)|^2$ in turn, stays robust and is shifted to lower values of the threshold for $\Delta_\text{Z}$ [see App.~\ref{sec.:App-Coefficients}, Fig.~\ref{fig.:coeffRe} and Eq.~\eqref{eq.:zeroenergyScattering}]. Conclusively, while the states for $\Delta_\text{Z}=0.1\Delta$ and $\Delta_\text{Z}=0.2\Delta$ are strongly flux dependent, the zero energy states at $\Delta_\text{Z}=0.5\Delta$ remains unaffected. This is the main result of this paper and constitues a clear signature to discriminate between the presence of ABS and MBS.

\section{Summary and Conclusion}\label{sec.:sectionV}
In this article we calculated WTDs for a superconducting hybrid setup on quantum spin hall edge states, proposed by Fu and Kane~\cite{fu2009}. First we presented the energy dependence of the scattering events appearing in the Josephson junction without magnetic field, from which we conclude that the distribution of the WTD for local Andreev reflection is independent of the size of the voltage window within the SC gap $\Delta$, since the resonant ABS, at which the reflection becomes zero can be neglected and the corresponding results can be seen as a perfectly transmitting channel following Wigner-Dyson statistics. For the second scattering process, the electron transmission, the number of included resonances within the voltage window induces specific oscillations in the WTD. For a large number of resonances, the WTD leads to a Poisson distribution. Analytically, we verified that the coherence of sequentially aligned Lorenzian shaped resonances can be used within the formalism of Brandes~\cite{Brandes2008} to explain the oscillations in the respective WTD. Most importantly, we increased a Zeeman field in the area between the two superconducting islands, such that the transition in the WTD from ABSs to MBSs was studied in great detail. We have shown in the absence of the superconducting phase difference that the WTD along the topological phase transition reflects a coherently oscillating dependence to a purely singe channel dependence, which clearly distinguishes between the two situations. Interestingly, when the phase difference reaches $\pi$, all ABS laying at energies greater than zero almost fully prohibits normal electron reflection, while the state at zero energy stays robust. Conclusively a transition in the WTD is directly visible, due to a reduced threshold of the Zeeman field.  
We would like to conclude with a practical application of our setup. Experimental setups of topological Josephson junctions already have been studied in the literature. Depending on the material realization of the two dimensional TI different SCs have been used. If the WTD is not yet a routinely measured quantity, many progresses toward single electron measurements are promising \cite{Singleelectron1,Singleelectron2,Singleelectron3,Singleelectron4,Singleelectron5,Singleelectron6,Singleelectron7,Basset2012} and the WTD has been already measured in several experiments \cite{Singleelectron5,Flindt2009,Ubbelohde2012}.

\appendix

\section{Scattering Coefficients}\label{sec.:App-Coefficients}
\begin{figure*}
\includegraphics[width=0.7500\linewidth]{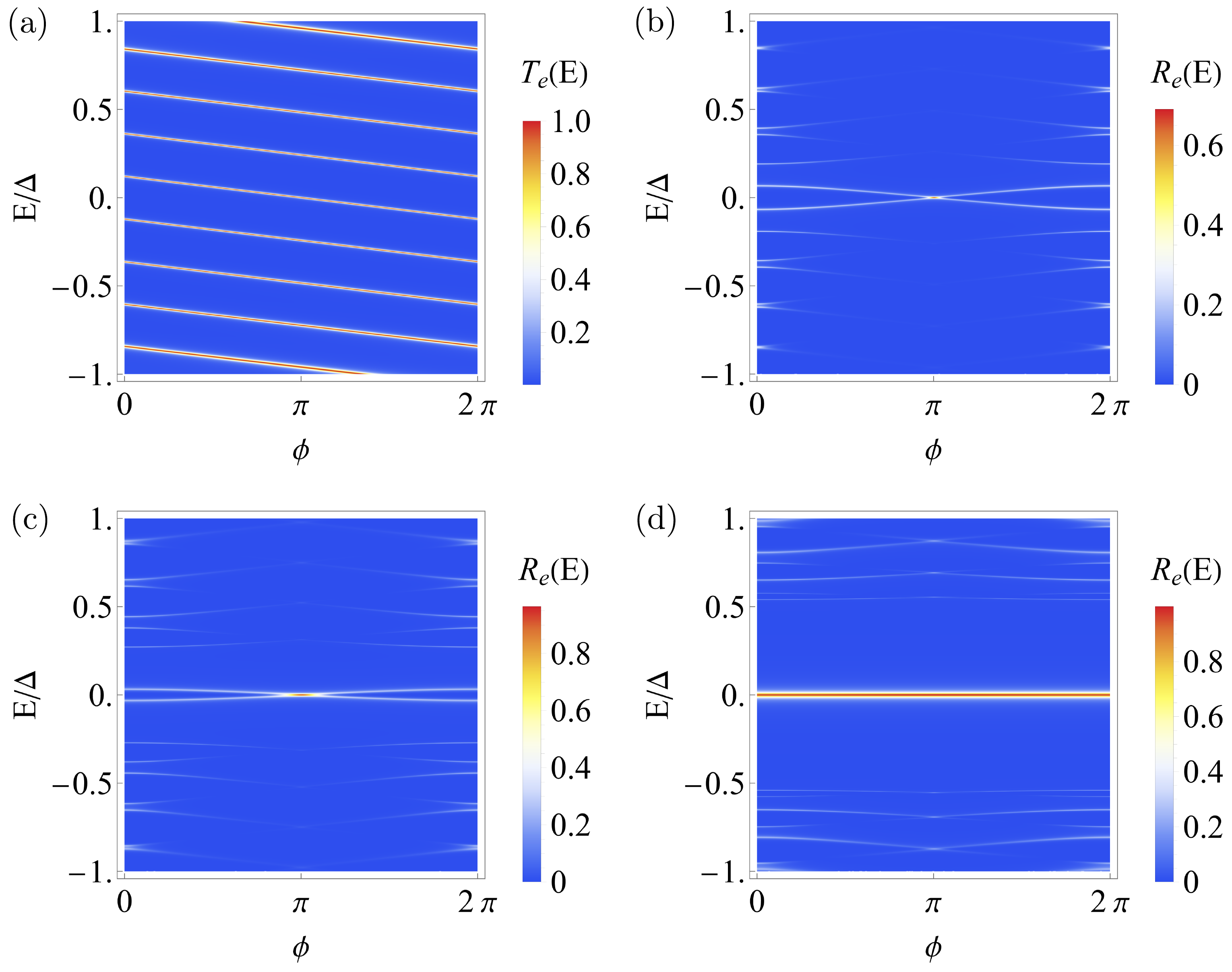}
\caption{Phase dependence of the scattering coefficient (a) $T_e(E)$ at $\Delta_\text{Z}=0.0\Delta$ , (b) $R_e(E)$ at $\Delta_\text{Z}=0.1\Delta$, (c) $R_e(E)$ at $\Delta_\text{Z}=0.2\Delta$ and (d) $R_e(E)$ at $\Delta_\text{Z}=0.5\Delta$. The parameters are the same as in Fig.~\ref{fig.:coef1}.}
\label{fig.:coeff2}
\end{figure*}

In this section we explain details of the BTK formalism for the calculations of the scattering coefficients. In general, we consider a N-S-F-S-N -junction as explained in the main text. The full scattering state is composed out of five parts, given by
\begin{eqnarray}\label{eq.:14}
\psi_\text{I}(x) && =  \vec{\mathbf{e}}_1e^{ikx}+r_{e}\vec{\mathbf{e}}_2e^{-ikx}+r_{h}\vec{\mathbf{e}}_4e^{ikx}\text{,}\nonumber\\
\psi_{\text{II}}(x)  &&=\sum_{i=1}^4a_i\vec{\mathbf{u}}_ie^{ik_ix}\text{,}\nonumber \\
 \psi_{\text{III}}(x) &&=\sum_{i=1}^4b_i\vec{\mathbf{v}}_ie^{ik'_ix}\text{,}\nonumber \\
 \psi_{\text{IV}}(x) &&=\sum_{i=1}^4c_i\vec{\mathbf{w}}_ie^{ik_ix}\text{,}\nonumber\\
 \psi_{\text{V}}(x) &&=t_{e}\vec{\mathbf{e}}_1e^{ikx}+t_{h}\vec{\mathbf{e}}_3e^{-ikx} \text{,}
\end{eqnarray}
where the eigenstates ($\vec{\mathbf{u}}_i,\vec{\mathbf{v}}_i,\vec{\mathbf{w}}_i$) are those of the Hamiltonian for the corresponding region and the vectors $\vec{\mathbf{e}}_i$ in the outer normal N-regions are the 4-dimensional Euclidean basis vectors of Eq.\eqref{eq.:1}. The continuity conditions, according to the setup, must hold and allow us to calculate all coefficients of the scattering states, especially the transmission
and reflection coefficients. The explicit conditions read
\begin{eqnarray} \label{eq:15}
\psi_\text{I}(0)&&=\psi_{\text{II}}(0)\textit{,}\nonumber \\
\psi_{\text{II}}(L)&&=\psi_{\text{III}}(L)\textit{,}\nonumber \\
\psi_\text{III}(L+L_m)&&=\psi_{\text{IV}}(L+L_m)\textit{,}\nonumber \\
\psi_\text{IV}(2L+L_m)&&=\psi_{\text{V}}(2L+L_m) \textit{,} 
\end{eqnarray}
where the two outer wave-functions (I and V) are those of the bare edge states, containing the necessary reflection (in $\psi_\text{I}$) and transmission (in $\psi_\text{V}$) coefficients. We have chosen the structure in such a way that the superconducting regions II and IV both have the length $L$, while the magnetic region III has length $L_m$.
 In general the scattering coefficients are complex. Without Zeeman field we get the expression
\begin{eqnarray}\label{eq.:16}
r_{h}&&=\frac{\left(-1+e^{2 i L \Omega _{\Delta }}\right) \left(-1+e^{i \left(2 E  L_m+\phi \right)}\right) \left(-v^2+u^2 e^{2 i L \Omega _{\Delta }}\right)}{D}\text{,}\nonumber \\
t_{e}&&=\frac{\left(u^2-v^2\right)^2 e^{-2 i L \left(E -\Omega _{\Delta }\right)}}{D}\text{,} \\
D&&=u^2 v^2 \left(-1+e^{2 i L \Omega _{\Delta }}\right){}^2 e^{i \left(2 E  L_m+\phi \right)}\nonumber\\
&&+\left(u^2 e^{2 i L \Omega _{\Delta }}-v^2\right) \left(u^2-v^2 e^{2 i L \Omega _{\Delta }}\right),\nonumber 
\end{eqnarray}
where $\Omega _{\Delta }=\sqrt{E^2-\Delta^2}$ and $ u^2(v^2)=\Big(E \pm \sqrt{E ^2-\Delta ^2}\Big)/\Delta$. For the more complex situation of a non zero Zeeman field in the middle region, the results are to cumbersome to present, such that we present the important coefficients $T_e(E)=|t_e|^2$ and $R_e(E)=|r_e|^2$ in Fig.~\ref{fig.:coeff2}. At zero energy the scattering coefficients can be simplified to
\begin{eqnarray}\label{eq.:zeroenergyScattering}
r_{e}&&=-\frac{2 i \sinh \left(2 L_m \Delta _Z\right)}{D}\text{,}\nonumber\\
r_{h}&&=\frac{2 \sin (\phi ) \sinh (2 \Delta  L)-i (\cos (\phi )+1) \sinh (4 \Delta  L)}{D}\text{,}\nonumber \\
t_{e}&&=\frac{4 \cosh \left(L_m \Delta _Z\right) \left(\cosh ^2(\Delta  L)+e^{-i \phi } \sinh ^2(\Delta  L)\right))}{D}\text{,}\nonumber \\
t_{h}&&=\frac{2 \left(1+e^{i \phi }\right) \sinh (2 \Delta  L) \sinh \left(L_m \Delta _Z\right)}{D}\text{,}\nonumber \\
D&&=  \cosh (4 \Delta  L)[1+\cos (\phi )]+2 \cosh \left(2 L_m \Delta _Z\right)\nonumber\\
&&+[1-\cos (\phi )]\text{.}
\end{eqnarray}

\begin{figure}
\includegraphics[width=0.800\linewidth]{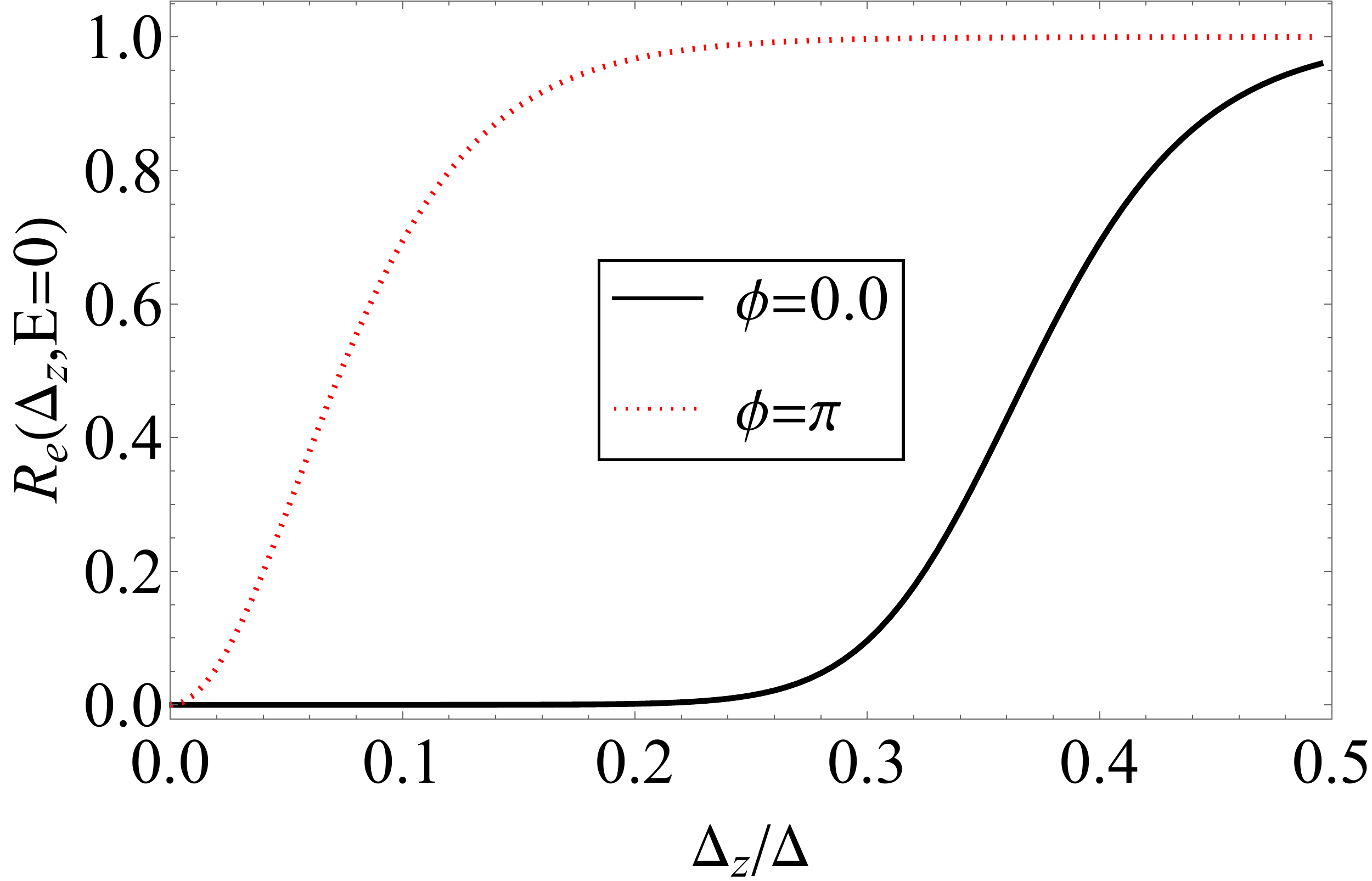}
\caption{Scattering probability of normal electron reflection at zero energy for $\phi=0$ and $\phi=\pi$. The parameters are the same as in Fig.~\ref{fig.:coef1}.}
\label{fig.:coeffRe}
\end{figure}

\section{Rate equation single dot}\label{sec.:App-RateeqSD}
We present in the appendix the calculations for the rate equation for a double barrier following the results of~\citep{Brandes2008}. Starting with the splitting of the Liouvillian
\begin{eqnarray}
\mathcal{L}= \begin{pmatrix}
-\Gamma_1 &0 \\
\Gamma_1 & -\Gamma_2 
\end{pmatrix}+ \begin{pmatrix}
0 & \Gamma_2 \\
0 & 0 
\end{pmatrix}=\mathcal{L}_0+\mathcal{J},
\end{eqnarray}
where $\mathcal{J}= |2\rangle\langle \bar{2}|$, with $\langle \bar{2}|=(0,\Gamma_2)$ and $|2\rangle=(1,0)^\top$, measuring jumps from state $2$ to $1$. The corresponding WTD is given by
\begin{equation}
\hat{W}(z)=\langle \bar{2}|(z-L_0)^{-1}|2\rangle=\frac{\Gamma_1}{(z+\Gamma_1)}\frac{\Gamma_2}{(z+\Gamma_2)},
\end{equation}
such that we get with the inverse Laplace transformation $F(z)=1/(z+a)\rightarrow f(\tau)=e^{-a\tau}$,
\begin{equation}
W(\tau)=\frac{\Gamma_1\Gamma_2}{(\Gamma_1-\Gamma_2)}\left(e^{-\Gamma_2\tau}-e^{-\Gamma_1\tau}\right)=\Gamma^2 \tau e^{-\Gamma\tau},
\end{equation}
while we assumed in the last step $\Gamma_1=\Gamma_2=\Gamma$ to derive the equation used in the main text of the article. From
\begin{equation}
\frac{d^nW(\tau)}{d\tau^n}=(-1)^n e^{-\Gamma  \tau } \Gamma ^{n+1} (\Gamma  \tau -n),
\end{equation}
 the maximum of $W(\tau)$ is at $\tau=1/\Gamma$.

\section{Waiting time distribution for series of Lorenzian resonances}\label{sec.:App-Rateeq.}
\begin{figure}[h]
  \includegraphics[width=0.900\linewidth]{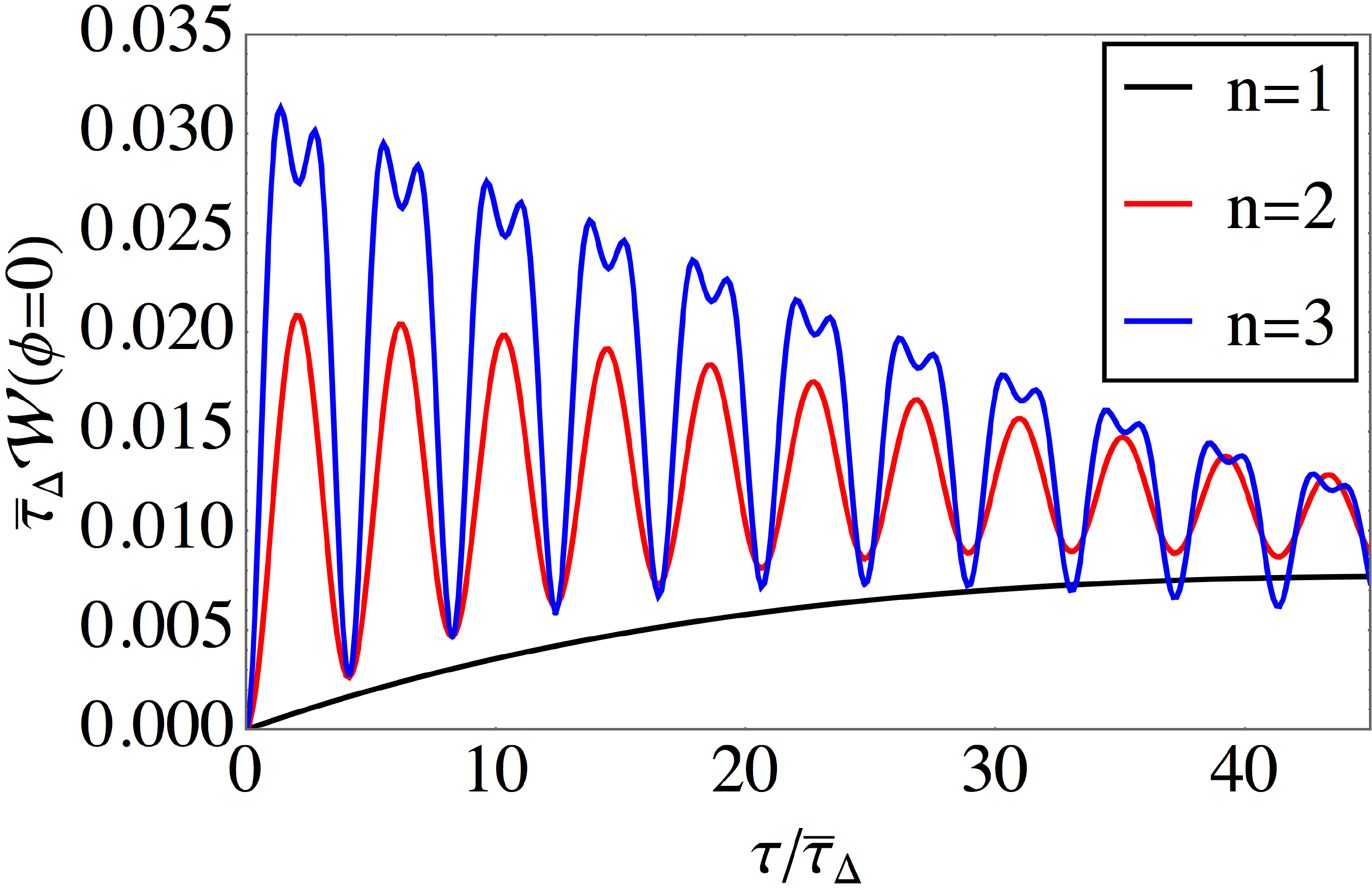}
  \caption{(Color online) Waiting Time Distributions for several numbers of resonance states. The parameters are given by the FWHM of the transmission coefficients for $\Gamma=0.021\Delta$ and $\omega=0.24\Delta 2\pi$.}
  \label{fig.:WTDanaly}
\end{figure}

In this appendix we explain the derivation of the WTD for the inclusion of an arbitrary number of Lorenzian shaped resonances. As explained in the main text, by including more then one state in the voltage window one finds an oscillatory behavior within the resulting WTD. In general, the distribution extend to waiting times much larger than $h/eV$, the typical time scale at which correlations between electrons or holes are important. We will then assume that the stochastic process can be reduced to a renewal process \cite{WTD2}. In that case, the WTD is related to the second order correlation function or equivalently to the two-body density matrix of the field. In order to compute the second order correlation function, we assume a wave function of the form,
\begin{equation}
\psi_n^\dagger(x)=e^{ikx}a_{k,n}^\dagger,
\end{equation}
such that the resulting one body density matrix $\rho_{1,n}$ is given by 
\begin{eqnarray}
\rho_{1,n}(x,y)&&=\langle \psi_n^\dagger(x)\psi_n^{}(y)\rangle=\int dkdk' e^{i(kx-k'y)}\langle a_{k,n}^\dagger a_{k',n}^{} \rangle \nonumber\\
&&=\int dkdk' e^{i(kx-k'y)}\delta_{k,k'}f(k)\big[\sum_{j=1}^{n}\frac{\gamma^2}{(k-j k_0)^2+\gamma^2}\big]\nonumber \\
&&=\int dk e^{ik(x-y)}\big[\sum_{j=1}^{n}\frac{\gamma^2}{(k-j k_0)^2+\gamma^2}\big]\nonumber\\
&&=\pi\gamma e^{-\gamma |x-y|} \big[\sum_{j=1}^{n}e^{ik_0j(x-y)}\big],
\end{eqnarray}
where we assumed that the propagator is given by the sum of Lorenzian shaped distributions with a maximum of unity, an equidistant separation of $k_0$ and $\gamma$ being half of the Full width at half maximum. We assumed that these states are perfectly in the voltage window, such that the Fermi distribution $f(k)$ is unity. Next, we evaluate the two body density matrix by using Wick's theorem, or equivalently by taking advantage of the fact that the field is described by a determinantal process. This gives
\begin{eqnarray}
\rho_{2,n}(x,y)&=&\left|
\left(
\begin{array}{cc}
 \rho_{1,n}(x,x) & \rho_{1,n}(y,x) \\
 \rho_{1,n}(x,y) & \rho_{1,n}(y,y) \\
\end{array}
\right)\right|\\
&=&(\pi\gamma)^2\left( n^2-e^{-2\gamma |x-y|}\left|\sum_{j=1}^{n}e^{ik_0j(x-y)}\right|^2\right)\nonumber.
\end{eqnarray}
Going to the time domain by assuming a linear dispersion relation, we simply get 
\begin{equation}
\tilde{\rho}_{2,n}(t,t')=(\pi \Gamma)^2\left( n^2-e^{-2\Gamma |t-t'|}\left|\sum_{j=1}^{n}e^{iv_Fk_0j( t-t')}\right|^2\right),
\end{equation}
with the effective width $\Gamma=v_F\gamma$.  The corresponding relaxation current \cite{Brandes2008} is then given by
\begin{eqnarray}
  I_n(t,t')&=&I\frac{\tilde{\rho}_{2,n}(t,t')}{\tilde{\rho}_{2,n}(0,\infty)}\\
  &=& n\frac{\Gamma}{2}\left( 1-\frac{ e^{-2\Gamma |t-t'|}}{n^2}\left|\sum_{j=1}^{n}e^{iv_Fk_0j( t-t')}\right|^2\right)\nonumber.
\end{eqnarray}
Since the process is stationary we may set $t=0$ and $t'=t$. We then rewrite the absolute value of the sum of exponentials as 
\begin{equation}
\left|\sum_{j=1}^{n}e^{iv_Fk_0j(-t)} \right|^2 =n+\sum_{j=1}^{n} 2 (j-1) \cos [v_Fk_0t (n+1-j)],
\end{equation}
such that we have for example,
\begin{eqnarray}
I_1(t)&&=\frac{\Gamma}{2}\left( 1-e^{-2\Gamma |t|}\right)\\
I_2(t)&&=\Gamma\left( 1-\frac{ e^{-2\Gamma |t|}}{2}\big[1+\cos (v_Fk_0t)\big]\right).
\end{eqnarray}
This further leads to three different types of standard Laplace transforms, namely
\begin{eqnarray}
&&\int\limits_{0}^{\infty}dte^{-zt}= \frac{1}{z},\\
&&\int\limits_{0}^{\infty}dte^{-zt}\frac{ e^{-2\Gamma t}}{n}=\frac{1}{n(z+2\Gamma)},\\
  &&\int\limits_{0}^{\infty}dte^{-zt}\frac{ e^{-2\Gamma t}}{n^2}2 (j-1) \cos [v_Fk_0t (n+1-j)]\\
  &&=\frac{2 (j-1)(z+2\Gamma)}{n^2[(z+2\Gamma)^2+(\omega_j)^2}\nonumber,
\end{eqnarray}
 with $\omega_j=v_Fk_0 (n+1-j)$, it transforms the current to
\begin{equation}
I_n(z)=n\frac{\Gamma}{2}\left[\frac{1}{z}-\frac{1}{n(z+2\Gamma)}-\sum_{j=1}^{n}\frac{2 (j-1)(z+2\Gamma)}{n^2[(z+2\Gamma)^2+(\omega_j)^2}\right].
\end{equation}
To get the waiting time distribution, we assume a renewal process, which is justified if the typical waiting time is much larger than $\bar{\tau}=h/eV$. Following Ref.~\cite{Brandes2008}, we calculate
\begin{equation}
w_n(z)=\frac{I_n(z)}{1+I_n(z)}.
\end{equation}
The inverse Laplace transform is for $n>1$ already difficult to handle. For example, the $n=2$ WTD in Laplace space reads
\begin{equation}
w_2(z)=\frac{\Gamma  \left(\omega^2 (4 \Gamma +z)+4 \Gamma  (2 \Gamma +z)^2\right)}{\omega^2 \left(4 \Gamma ^2+2 z^2+5 \Gamma  z\right)+2 (2 \Gamma +z)^2 \left(2 \Gamma ^2+z^2+2 \Gamma  z\right)},
\end{equation}
with the inverse Laplace transformation
\begin{equation}
w_2(t)=\frac{1}{2\pi i}\int\limits_{\alpha-i\epsilon}^{\alpha+i\epsilon} dz e^{tz}\frac{\Gamma  \left(\omega^2 (4 \Gamma +z)+4 \Gamma  (2 \Gamma +z)^2\right)/2}{(z-z_1)(z-z_1^*)(z-z_2)(z-z_2^*)},
\end{equation}
with $\alpha$ a real number defining the so called Bromwich contour. To calculate the contour, we evaluate the poles numerically. The structure of the poles is given by $z_1$, being close to the real axis containing a small imaginary part and result out of the term of the numerator with the highest power of $\omega$ ($\omega\gg \Gamma$). Further, depending on the number of resonances the rest of the poles contain a large imaginary part resulting in oscillations.  We present in Fig.~\ref{fig.:WTDanaly} the analytic results for the first three resonances. 

Besides the great agreement to the numerical results, we find a difference in the damping of the oscilations, which might be accounted by deviations from the Lorenzian shape at higher energy within the model used in the main text [see.\ref{fig.:coef1} (a)].

\end{document}